\documentclass[useAMS,usenatbib]{mn2e}
\usepackage{graphicx}

\title[Polarimetric observations of NRAO 530 on pc scale]{Multi-waveband polarimetric
observations of \mbox{NRAO 530} on parsec-scale}
\author[Y. J. Chen, Z.-Q. Shen and S.-W. Feng]{Y. J.
Chen$^{1}$\thanks{E-mail: cyj@shao.ac.cn}, Z.-Q. Shen$^{1}$ and
S.-W. Feng$^{2}$\\
$^{1}$Key Laboratory for Research in Galaxies and Cosmology,
Shanghai Astronomical Observatory, Chinese Academy of
Sciences, \\Shanghai 200030, P.R. China\\
$^{2}$Department of Space Science and Applied Physics, Shandong
University at Weihai, Weihai 264209, China}
\begin{document}

\date{Accepted 2010 June 12. Received 2010 June 11; in original form 2009 October 18}

\pagerange{\pageref{firstpage}--\pageref{lastpage}} \pubyear{2002}

\maketitle

\label{firstpage}

\begin{abstract}
We report on VLBA polarimetric observations of NRAO 530 at 5, 8, 15,
22, and 43 GHz within an interval of one week in February 1997.
Total intensity, fractional polarization and electric vector
position angle (EVPA) distributions at all these frequencies are
presented. Model fitting has been performed to the full polarization
visibility data, from which the fitted southmost component \emph{A}
is confirmed as the core of the radio structure with relatively high
brightness temperature and hard spectrum between 15 and 43 GHz in
comparison with the central component \emph{B} of dominant flux. The
relatively high degree of polarization for the component \emph{A}
may arise from its complex radio structure, which is resolvable at
86 GHz. As a contrast, the component \emph{B} shows a well fitted
power-law spectrum with the spectral index of about $-0.5$
($f\propto\nu^\alpha$), and a linear correlation between EVPAs and
wavelength square with the observed rotation measure (RM) of about
$-1062~ \rm{rad~m^{-2}}$, indicating its structural singleness.
Assuming that the component \emph{B} has a comparable degree of
polarization without depolarization at these frequencies, the
decrease in fractional polarization with wavelength mainly results
from opacity and Faraday rotation, in which the opacity plays quite
a large part of role. A spine-sheath like structure in fractional
polarization ($m$) is detected covering almost the whole emission
region at 5 and 8 GHz, with degree of polarization relatively low
along the jet spine, and getting higher toward two sides of the jet.
The linear polarization at 5 GHz shows 3 separate polarized emission
regions with alternately aligned and orthogonal polarization vectors
down the jet. The polarization goes to zero between the top two
regions, with the highest polarization level occurring at the top
and bottom. The 5 and 8 GHz images show EVPA changes across the
width of the jet as well as along the jet. These complex
polarimetric properties can be explained in terms of either the
presence of a large helical magnetic field or tangled magnetic
fields compressed and sheared down the jet, which can be further
determined by multi-frequency polarimetric VLBI observation with
sufficient high resolution and sensitivity spanning over an
appropriate frequency range.

\end{abstract}


\section{Introduction}
In the leading model for jet production, acceleration, and
collimation in blazars, magnetic field plays an important role (e.g.
\citealt{mei01, mck06}). The poloidal magnetic fields are supposed
to be wound up by the differential rotation of a rotating disk or
ergosphere surrounding a central suppermassive black hole,
propagating outward in the polar directions with a tight helical
pattern. On pc scale or beyond the jet acceleration region, the
magnetic fields within jet might maintain a tight helical pattern
\citep{lyu05}, become chaotic, or possibly get compressed and
sheared \citep{jos07}. The degree of order and geometry of the
magnetic field will differ from case to case, and therefore can help
to better understand the physical conditions in a jet.

Since the jet emission is mainly synchrotron radiative, and hence
linearly polarized with electric vector perpendicular to the
projection on the sky of the magnetic field \citep{beg84}, the
magnetic geometry and order of degree can be revealed to some degree
through polarimetric Very Long Baseline Interferometry (VLBI)
observation. The dominant transverse magnetic field is often
ascribed to shock compression, and the dominant longitudinal one to
the effect of shear or interaction with surrounding medium (e.g.
\citealt{lai80, hug89}). However, both cases can also be interpreted
in terms of intrinsic helical magnetic fields, which appear more
natural and simpler ( e.g. \citealt{gab04}). It is difficult to
distinguish between transverse magnetic fields due to a toroidal
field component and due to shock compression. Under the
circumstances, the measurement of RM gradient across the jet is
proposed to test the magnetic helicity within or wrapping around the
jet \citep{bla93}, which has been detected in some sources such as
3C 273 \citep{asa02,zav05} through multi-band polarimetric VLBI
observation with sufficiently good visibility data.

As a typical blazar, NRAO 530 (J1733-1304) is strong and variable in
almost the whole wave bands from radio to $\gamma$-ray. Long term
monitoring at cm wavelengths from 1967 to 2003 shows a bright
outburst peaking around 1997 \citep{pya06}, which is almost
coincident with our VLBI observations. In this paper, we present the
polarimetric VLBI observational results of \mbox{NRAO 530} at 5
frequencies. Model fitting has been done to the full polarization
data, with the physical properties of the southmost component
\emph{A} and the central component \emph{B} analyzed and discussed
in more detail. Since the total intensity structures have been
reported in \citet{fen06}, we here focus more on the polarimetric
properties of the source on parsec scale.

In \S2 we present the whole process of data reduction. Results are
shown in \S3, followed by discussions in \S4. A summary is given in
\S5. Throughout this paper, we take the cosmological parameters of
$H_0 = 71~\rm{km~s^{-1}~Mpc^{-1}}$, $\Omega_M = 0.27$, and $\Omega_
\Lambda = 0.73$ in calculating angular distance, with 1 mas of
angular size corresponding to 7.8 pc for \mbox{NRAO 530} with a
redshift of 0.902.

\section{OBSERVATIONS AND DATA REDUCTION}

We observed NRAO 530 at 5, 8, 15, 22, and 43 GHz in February 1997
with Very Long Baseline Array (VLBA) plus one VLA antenna. Three
separate observations were made within an interval of 7 days, with
the first observation at 5 and 8 GHz made on February 7, the second
one at 15 and 22 GHz on February 12, and the third at 43 GHz on
February 14. Some other specific observational information can be
found in \citet{fen06}. Based on the UMRAO data during the
observational period, the largest flux density variation at 5, 8,
and 15 GHz is about 4\%, and the observations at 15, 22, and 43 GHz
were made within an interval of 2 days, whose total flux density
available at 15 GHz almost kept constant within the observational
error of about 2\% from February 8 to 10. This means that a 2 day
interval of observations almost has no impact on the following
analysis of the resultant spectral indices and RMs derived from the
high 3 frequency observations.

The data reduction and imaging were done in the AIPS and DIFMAP
packages respectively by using standard techniques, as described in
\citet{zav04}. Careful data editing was first done to flag out those
obviously bad data points, including all the visibility data to
antenna MK at 5 and 8 GHz, and those visibilities to BR at 15 and 22
GHz  due to failure of correlation, and some other points due to the
too low elevation angle. Standard amplitude and phase calibration
were performed subsequently. Here, we'd like to stress that we make
antenna gain(s) calibration with `CLCOR' to one or more antennas in
AIPS before imaging because at least one antenna gain is required to
be adjusted at all but 43 GHz.

NRAO 530 was observed as a calibrator during the observations. It is
very strong and compact, and hence the fringes can be easily
detected on almost all the baselines. Due to its relatively complex
radio structure, the task `CCEDT' in AIPS was used to separate the
total intensity distribution into several point sources by hand to
calibrate the instrumental polarization (`D-terms') of each antenna
with the AIPS task `LPCAL'. To ensure that the D-terms solutions
were acceptable, we checked the distribution of the normalized cross
visibilities in complex (real and imaginary) plane, with and without
instrumental calibration. We found that it gets well clustered after
the instrumental calibration is made. An additional effort is also
tried to obtain an independent set of D-terms from another
calibrator OV-236, which was observed for about half of the
on-source time on \mbox{NRAO 530}. These solutions are quite
consistent at 5 and 8 GHz. At 15 GHz and higher frequencies,there
are some differences. The distribution of the normalized
visibilities in complex plane gets better clustered, and the final
results have a better dynamic range, especially at 15 and 22 GHz
when using the D-terms from NRAO 530 alone. Therefor, in this paper,
we adopted the D-terms solutions derived from NRAO 530 alone.

The absolute EVPA calibrations at 5, 8, and 15 GHz were performed by
using the integral EVPA within 5 days of our observations from Radio
Astronomy Observatory Database at University of Michigan (UMRAO)
(cf. Aller et al. 1985). The absolute EVPAs are $52.8^\circ\pm
2.7^\circ$ at 5.0 GHz, $73.4^\circ\pm 2.2^\circ$ at 8.0 GHz, and
$70.8^\circ\pm 2.0^\circ$ at 15.4 GHz, respectively. At 22 and 43
GHz no absolute EVPAs are available, thus the true EVPAs at the two
frequencies are estimated by using RM of about
$-109~\rm{rad~m^{-2}}$ obtained at lower frequencies. The resultant
values at the two frequencies are about $75.5^\circ$ and
$76.6^\circ$, respectively. Corrections to EVPA are obtained by
doubling the difference between the integral EVPA from UMRAO and the
observed integral EVPA at each frequency, and then applied to the
visibility data with the AIPS task `CLCOR'. In comparison to the
UMRAO data, about 83\%, 86\% and 76\% of polarized flux at 5, 8 and
15 GHz get recovered respectively in our observations, that is to
say that the adopted EVPA corrections is roughly reasonable with the
corresponding uncertainties of $0.4^\circ$, $2.2^\circ$ and
$0.3^\circ$ due to the missing flux. Table~\ref{tb1} lists the
reference antenna and corresponding EVPA correction at each
frequency. The final images, including the total intensity,
fractional polarization, linear intensity and EVPA distributions are
shown in Figures 1 and 2, which are analyzed and discussed in \S3
and \S4, respectively.
\begin{table}
  \centering
  \begin{minipage}{80mm}
  \caption{Reference antenna and corresponding EVPA correction at each
  frequency}\label{tb1}
  \begin{tabular}{@{}lccccc@{}}
  \hline
  Obs. Freq. (GHz) & 5 & 8 & 15 & 22 & 43 \\ \hline
  Ref. Antenna & FD & FD & PT & PT & LA \\
$\chi$ (deg) & -33.6 & 156.6 & -16.4 & 19.1 & 77.6 \\
\hline
\end{tabular}
\end{minipage}
\end{table}

\section{RESULTS}
\subsection{Total intensity distribution and model fitting to the full visibility data}
\begin{figure}
\includegraphics[angle=-90,scale=0.5]{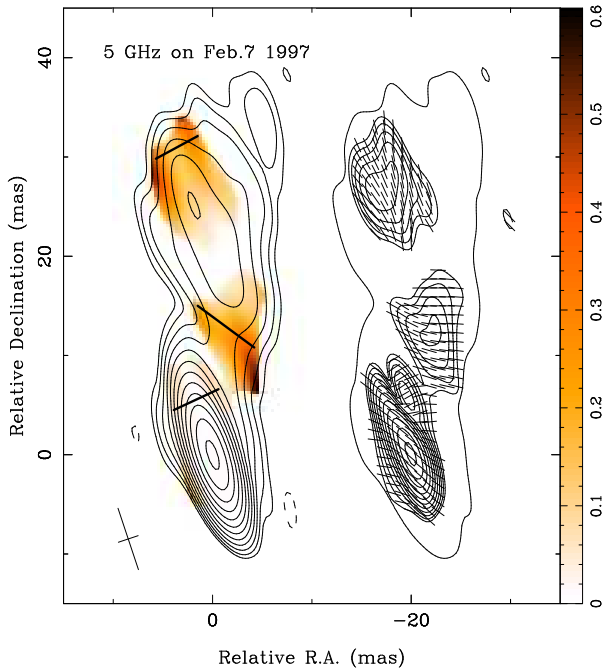}
\caption{VLBI polarimetric image of NRAO 530 at 5 GHz. Left:
fractional linear polarization image (color) overlaid by total
intensity contours. The restoring beam has dimensions of
$\rm{6.5~mas\times2.1~mas}$ at position angle $18.4^\circ$,
indicated at the bottom left corner. Contours start at
$2.9~\rm{mJy~beam^{-1}}$ and increase by factors of 2. The thick
lines indicate the location of the slices shown in Figure 5. Right:
EVPA distribution of NRAO 530 with contours of polarimetric
intensity overlaid. Contours start at $2.4~\rm{mJy~beam^{-1}}$ and
increase by a factor $\sqrt{2}$ for linearly polarized intensity
contours. \label{fig1}}
\end{figure}

\begin{figure*}.
\includegraphics[angle=-90,width=7.5cm, totalheight=7cm]{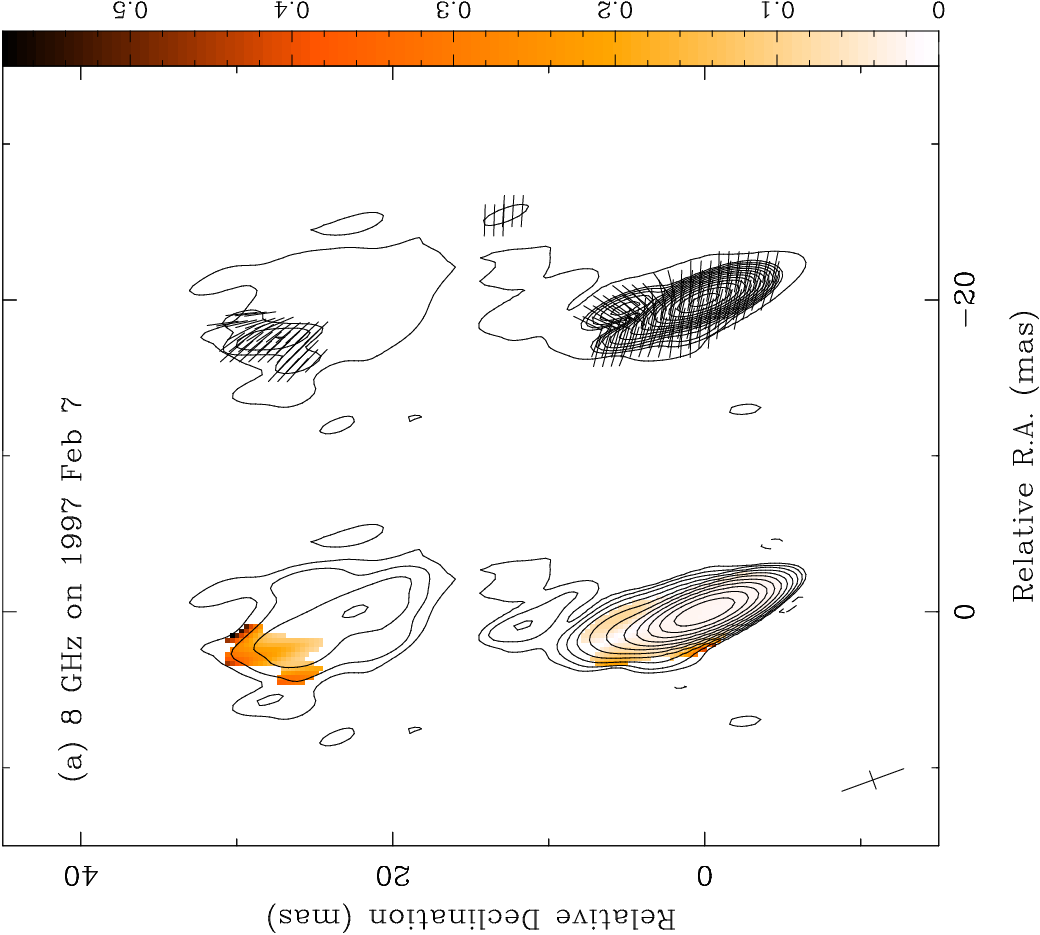}
\includegraphics[angle=-90,width=7.5cm, totalheight=7cm]{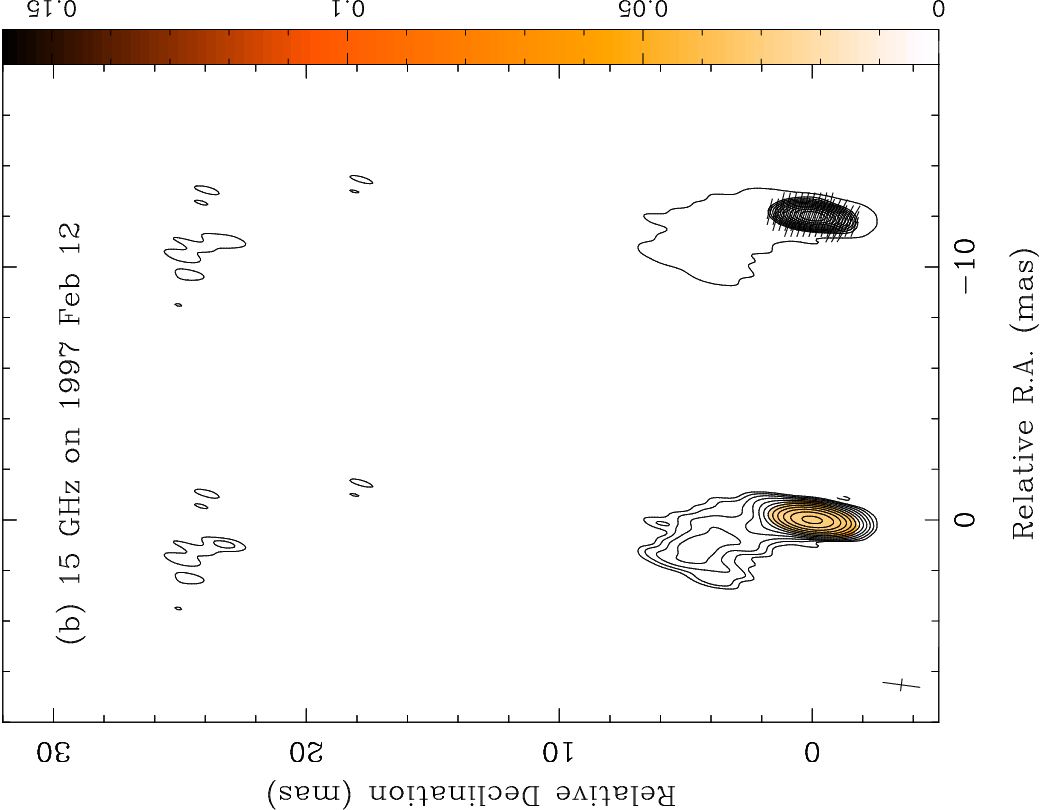}
\includegraphics[angle=-90,scale=0.4]{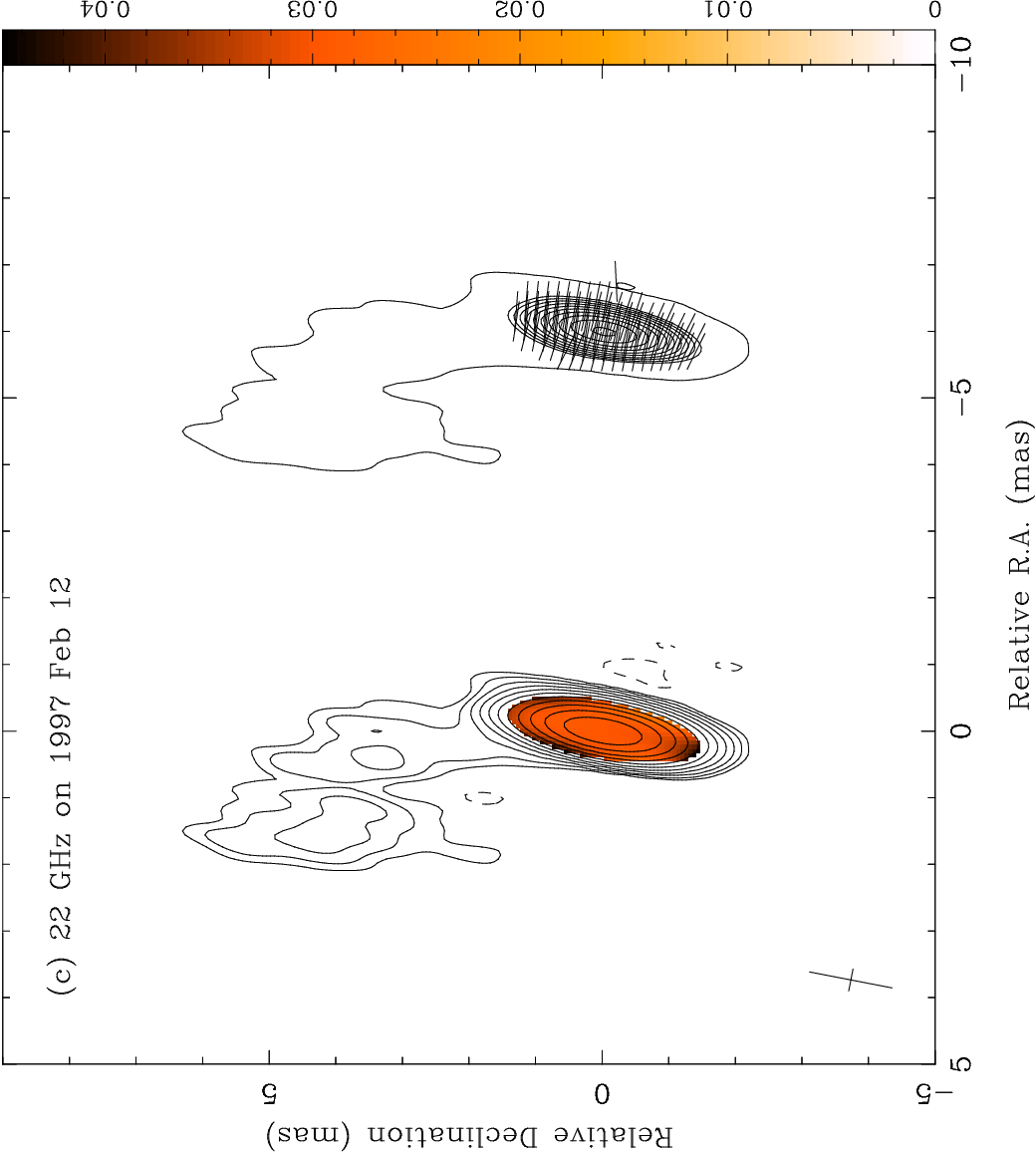}
\includegraphics[angle=-90,scale=0.4]{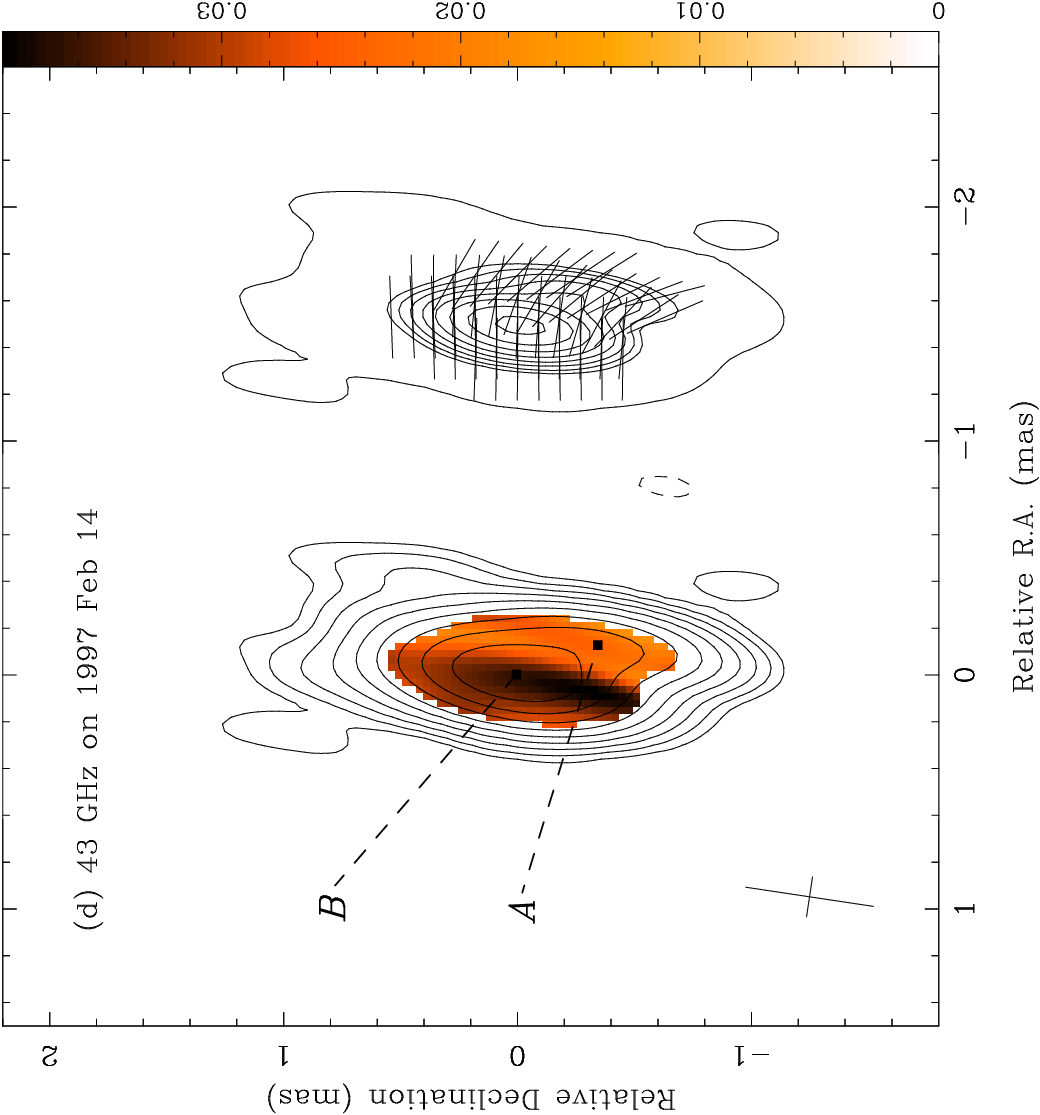}
\caption{The same as that in Fig. 1, but at four different
frequencies: (a) 8 GHz, (b) 15 GHz, (c) 22 GHz, and (d) 43 GHz. (a):
the restoring beam is $\rm{4.2~mas\times1.2~mas}$ at $20.0^\circ$,
contours start at $5.3~\rm{mJy~beam^{-1}}$ for total intensity and
$2.8~\rm{mJy~beam^{-1}}$ for linearly polarized intensity. (b): the
restoring beam is $1.5~\rm{mas\times0.5~mas}$ at $-7.7^\circ$,
contours start at $3.5~\rm{mJy~beam^{-1}}$ for total intensity and
$6.0~\rm{mJy~beam^{-1}}$ for linearly polarized intensity. (c): the
restoring beam is $\rm{1.3~mas\times0.3~mas}$ at $-10.9^\circ$,
contours start at $4.2~\rm{mJy~beam^{-1}}$ for total intensity and
$9.5~\rm{mJy~beam^{-1}}$ for linearly polarized intensity. (d): the
restoring beam is $\rm{0.55~mas\times0.17~mas}$ at of $-8.6^\circ$,
contours start at $7.5~\rm{mJy~beam^{-1}}$ for total intensity and
$10.5~\rm{mJy~beam^{-1}}$ for linearly polarized intensity. \emph{A}
and \emph{B} indicates the location of the fitted Gaussian
components \emph{A} and \emph{B} listed in Table 2. \label{fig2}}
\end{figure*}

The results and analysis of the total intensity distributions have
been presented in \citet{fen06}, where spectral fits to some of the
components and radiative mechanism are investigated in detail. Here,
we reproduce the intensity distributions at all the 5 frequencies as
contour profiles in the left panel of each map in Figures \ref{fig1}
and \ref{fig2}, with antenna gains calibrated to those antennas of
unreasonable magnitude by checking amplitude versus distance plots
at all but 43 GHz before imaging this time (e.g. \citealt{zav04}).
This gives rise to small differences in the resultant components'
amplitude in model fitting.
In terms of the recovered proportion of total flux density on pc
scale, the current results is more consistent in comparison to that
shown in MOJAVE data archive
(http://www.physics.purdue.edu/astro/mojave). The total intensity
distributions show that the detectable emission region decreases
rapidly with frequency. This is mainly due to large changes in
resolution, and the sensitivity difference at different frequencies
may also have contribution to it.

Model fitting is done using the MODELFIT program in the DIFMAP
package. Since a model fit to total intensity distribution is not
necessarily unique \citep{lis09}, we tried not to introduce any
artificial estimate into the process of model fitting to keep the
same components as presented in \citet{fen06}. We started with a
Gaussian component to model the Stokes \emph{I} structure. When the
reduced Chi-square no longer decreases and there are still
sufficient bright feature(s) present in the residual image, a new
Gaussian component is added into the residual image for further
fitting. The process was repeated until no apparently bright feature
could be found in the residual image. The linear polarization is
fitted using the same components as derived from the total intensity
fits with the only parameter flux left free to estimate the Stokes
\emph{Q} and \emph{U} flux density, and the corresponding linear
polarized flux and positional angle were obtained by using $P =
\sqrt{Q^2 + U^2}$ and $\chi = 0.5\arctan{U/Q}$, respectively.

The fitting results of the southmost component \emph{A} and the
component \emph{B} at phase tracking center (hereafter central
component) (see the right bottom panel of Figure 2) are shown in
Table~\ref{tb2}, where flux density ($f_\nu$), fractional
polarization ($m$), EVPA, major axis ($\theta_{maj}$), axial ratio
($\theta_{maj}/\theta_{min}$) and the derived brightness temperature
($T_B$) in the source frame for both components are listed in the
corresponding columns.
\begin{table*}
  \centering
  \caption{Model parameters to components \emph{A} and \emph{B}}\label{tb2}
  \begin{tabular}{@{}cccccccc@{}}
  \hline
  Component & Freq.  & $f_\nu$  & $m$  & $\chi$  &  $\theta_{maj}$ &
  $\theta_{min}/\theta_{maj}$ & $T_B$\\
    ID      & (GHz)  &  (Jy)     &  (\%)  & (deg)   &  (mas)   &       & ($10^{12}$K) \\ \hline
  \emph{A} & 15  & 1.01 & 4.97 & 49.9 & 0.046  & 1.0   & 6.86  \\
           & 22  & 1.74 & 5.45 & 56.0 & 0.052  & 1.0   & 4.30  \\
           & 43  & 1.30 & 5.22 & 21.8 & 0.035  & 1.0   & 1.87  \\
  \emph{B} & 5*  & 6.85 & 1.32 & 48.6 & 1.032  & 0.291 & 2.99  \\
           & 8*  & 9.76 & 2.01 & 73.9 & 0.615  & 0.214 & 5.73  \\
           & 15  & 8.49 & 2.95 & 71.6 & 0.184  & 0.586 & 6.11  \\
           & 22  & 7.30 & 3.13 & 83.7 & 0.269  & 0.467 & 1.47  \\
           & 43  & 4.86 & 4.46 & 91.8 & 0.169  & 0.893 & 0.34  \\
\hline
\end{tabular}
\end{table*}
The $T_B$ is estimated by using the following expression
\citep{ghi93},
\begin{equation}\label{eq1}
T_B=1.77\times10^{12}(\frac{f_\nu}{\rm{Jy}})(\frac{\nu}{\rm{GHz}})^{-2}(\frac{\theta_d}{\rm{mas}})^{-2}(1+z),
\end{equation}
where $f_\nu$ is flux density at frequency $\nu$, $\theta_d$ is
angular size $\theta_d = \sqrt{\theta_{maj}\theta_{min}}$ with
$\theta_{maj}$ and $\theta_{min}$ being major and minor axis,
respectively. Due to large difference in resolutions, we cannot
separate the southmost component \emph{A} at lower frequencies 5 and
8 GHz, the central component \emph{B} at the two frequencies
therefore probably contains a larger emission region in comparison
to that at higher frequencies, and will be investigated separately
from the other frequencies below.

\begin{figure}
\includegraphics[angle= 0,width=7cm,totalheight=9cm]{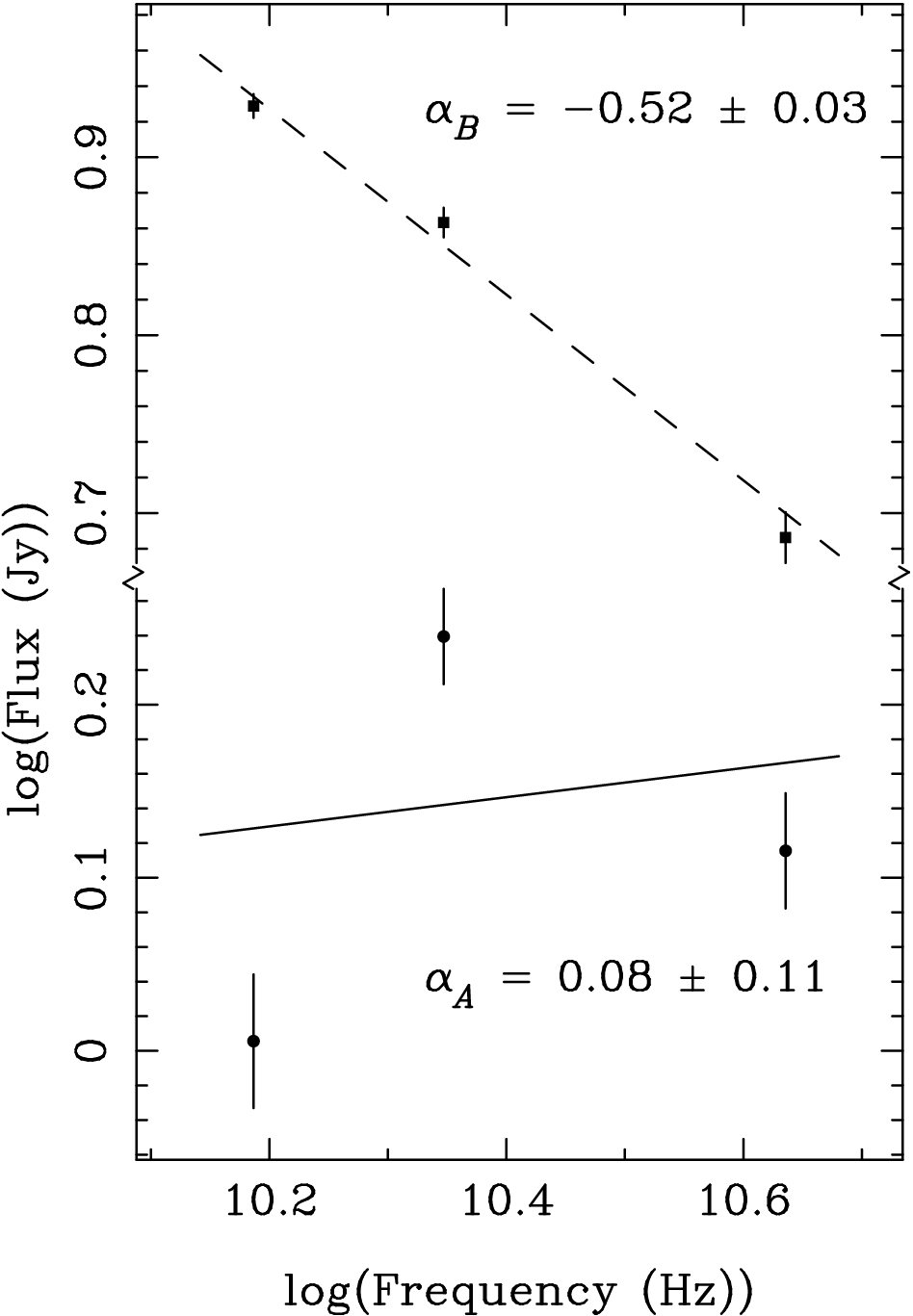}
\caption{Power-law spectral fitting to components \emph{A} (solid
line) and \textbf{B} (dashed line) from 15 GHz to 43 GHz.
\label{fig3}}
\end{figure}

The total intensity distributions show a typical core-jet structure
with jet roughly extending to north. Small fluctuations of jet
orientation occurs down the jet, which can be more clearly seen by
comparing radio structures at different frequencies. One can find in
Table~\ref{tb2} that the component \emph{A} lies in the extreme end
of the radio structure, and has relatively high brightness
temperature in comparison to the component \emph{B}, although its
flux density is relatively low. This may imply that it is the
component \emph{A} that represents the radio core of the source. To
have more evidence for the hypothesis, we fitted power law spectra
to both components from 15 GHz to 43 GHz, which is shown in
Figure~\ref{fig3}. The spectral indices for component \emph{A} and
\emph{B} are estimated to be $0.08\pm0.11$ and $-0.52\pm0.03$,
respectively. Obviously, the power-law spectrum for component
\emph{A} is fitted badly, but surely harder, while the component
\emph{B} exhibits a spectrum of quite a good power-law form, which
argues for the core hypothesis of the component \emph{A}, as
suggested in \citet{jos01} and \citet{fen06}.

\subsection{Fractional polarization distribution}

The polarization imaging results are shown in Figures \ref{fig1} and
\ref{fig2}, with fractional polarization superposed by the total
intensity contours,  and the EVPA structure superposed by linearly
polarized intensity contours at all the 5 frequencies. The 5 GHz
imaging result is shown alone for more discussions later. At this
frequency, the total and polarized intensity distribution have a
much better sensitivity in comparison with that at other frequencies
(see captions in Figures \ref{fig1} and \ref{fig2}). The lowest
contours for total and linearly polarized intensity distribution are
over 3 times the noise level ($\geq3\sigma$) at all the 5
frequencies. The emission region with detectable linear polarization
decreases gradually in size from $30~\rm{mas}$ to a few mas with
increasing observational frequency due to increasing resolution and
lower surface brightness of the more distant emission, making this
emission much harder to detect at high frequency.

As is shown in these images, the fractional polarization
distribution at all the 5 frequencies shows an overall tendency of
increase in degree of polarization with frequency. Such a frequency
dependent fractional polarization can be more clearly seen in Table
\ref{tb2}, where the degree of polarization for the component
\emph{A} available at 15, 22 and 43 Ghz and \emph{B} at all 5
frequencies are listed. The two components contain most of polarized
emission of the whole source, and hence may represent such a
dependence of polarization level on frequency. The overall
fractional polarization from low to high frequency is 1.6\%, 2.3\%,
2.9\%, 3.7\% and 2.9\%, respectively. At 43 GHz, the overall degree
of polarization is a little lower than that at 22 GHz. This may
mainly be due to the fact that the available radio structure at 43
GHz mostly lies in a region closer to the core, where there exists
more considerable substructures, and the opacity and Faraday
depolarization are more severe, thus dragging down the overall
polarization level (e.g. \citealt{tay98}). And the overall
dependence of polarization level on frequency is probably because
that at lower frequency, the depolarization by Faraday rotation is
more severe, and the resolution is relatively low.

In the outmost region of the 5 and 8 GHz fractional polarization,
there exists a prominent blob with extremely high fractional
polarization. In contrast, the core region shows a lower
polarization level although the absolute linear intensity is
relatively strong at all the 5 frequencies.  The relatively low
polarization level in the core region in comparison with that in jet
is also observed in many other sources such as 3C 273, 3C 279,
1803+784, etc, which may well be due to large opacity effect,
considerable substructures in the core region, and/or plasma in the
immediate vicinity of the AGN (e.g. \citealt{tay98, hom09}).

\subsubsection{Transverse fractional polarization}

\begin{figure}
\includegraphics[angle= 0, scale=0.35]{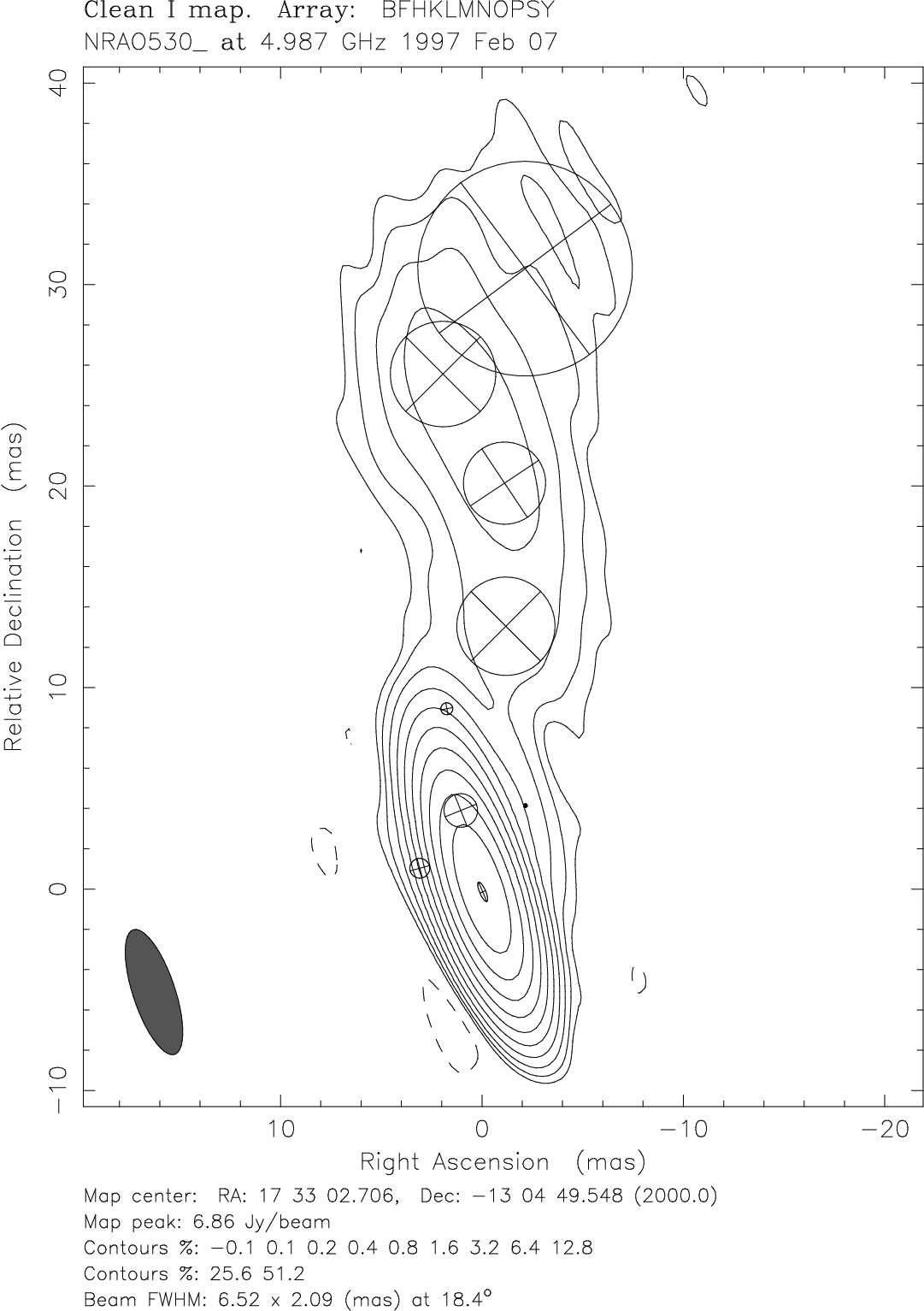}
\caption{Model fit to the total intensity visibilities at 5 GHz.
Each component is represented in an elliptic form with the fitted
position, size and orientation. \label{fig4}}
\end{figure}

\begin{figure*}
\includegraphics[angle= -90,scale=0.25 ]{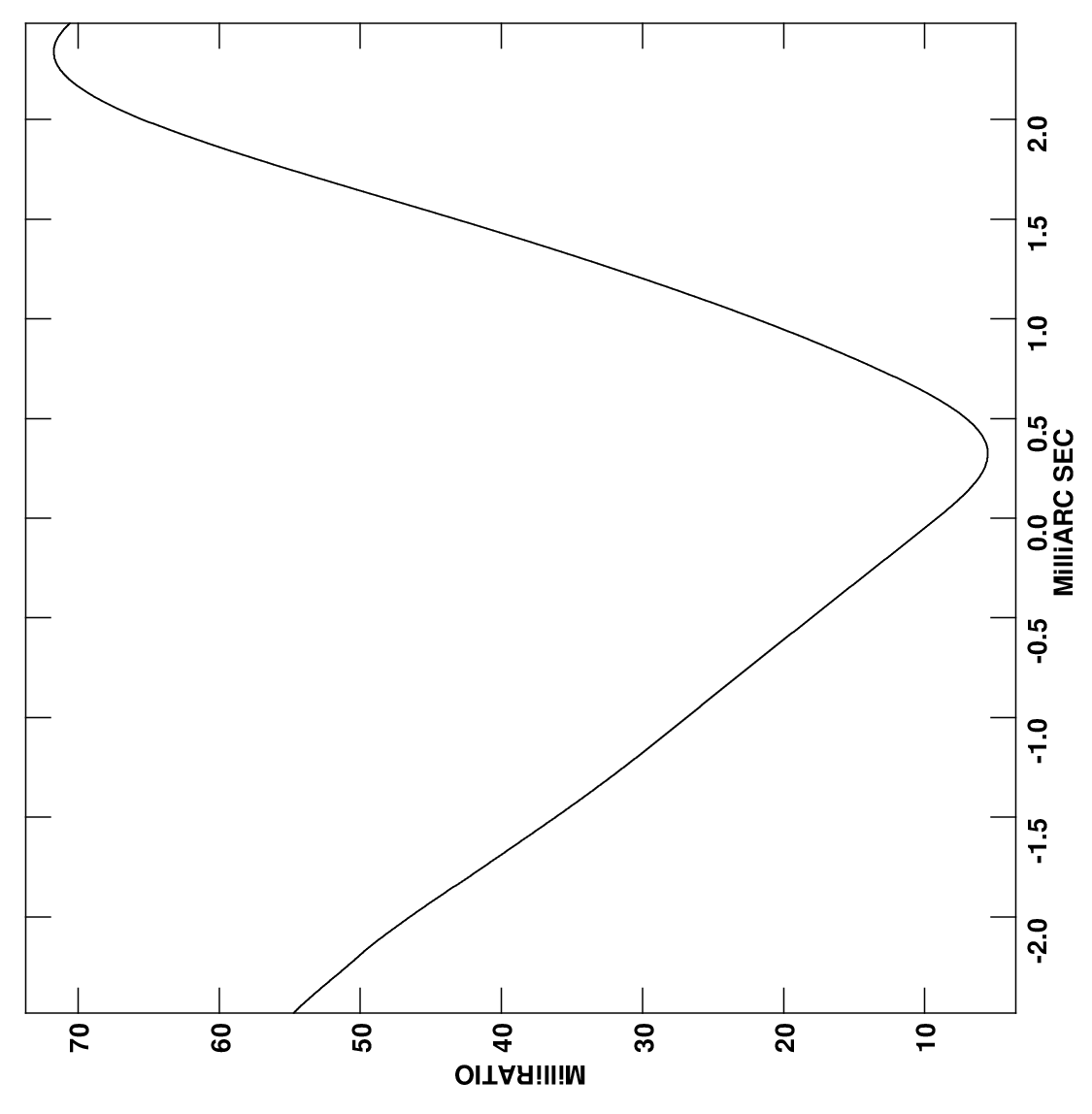}
\includegraphics[angle= -90,scale=0.25 ]{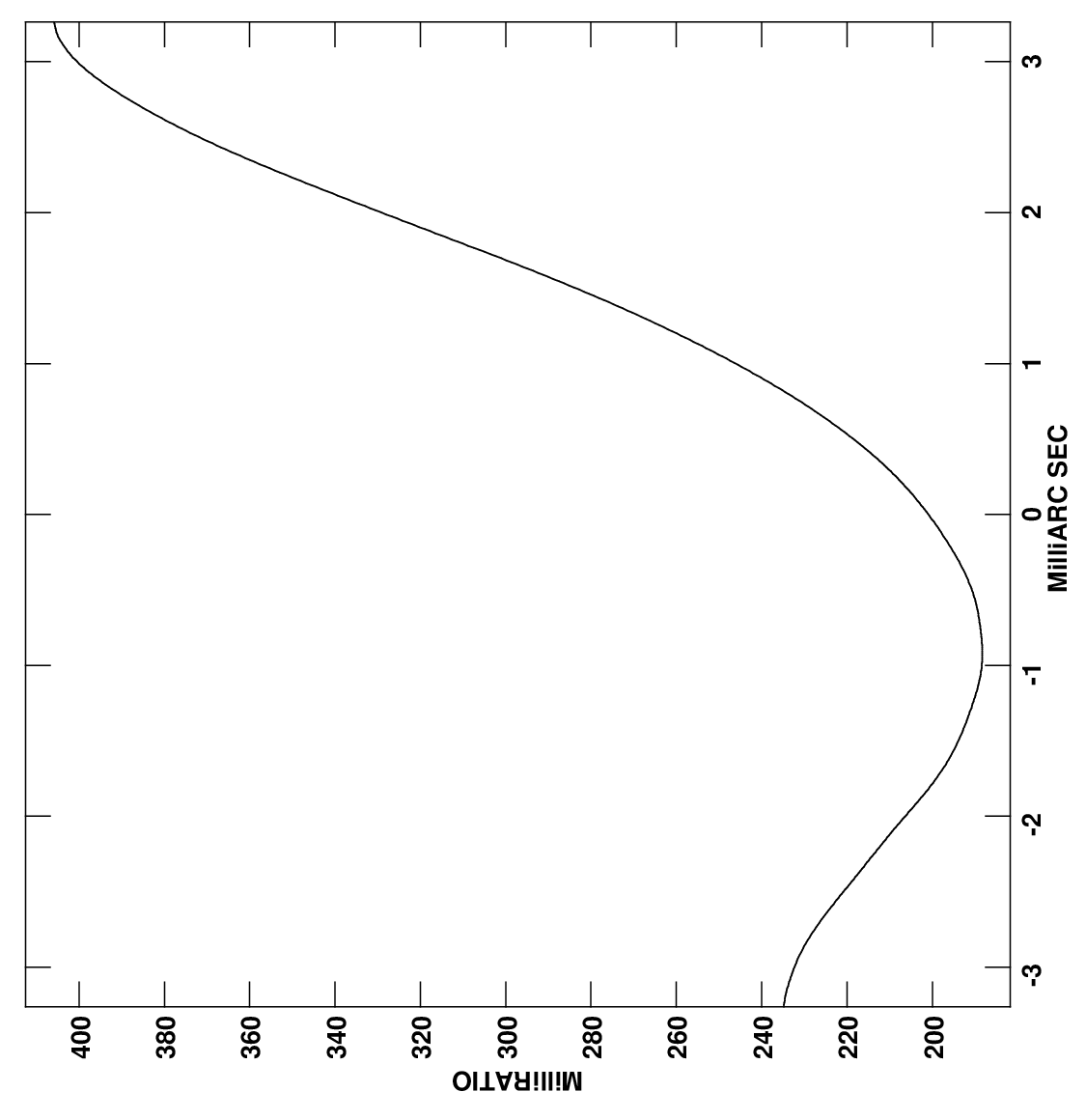}
\includegraphics[angle= -90,scale=0.25 ]{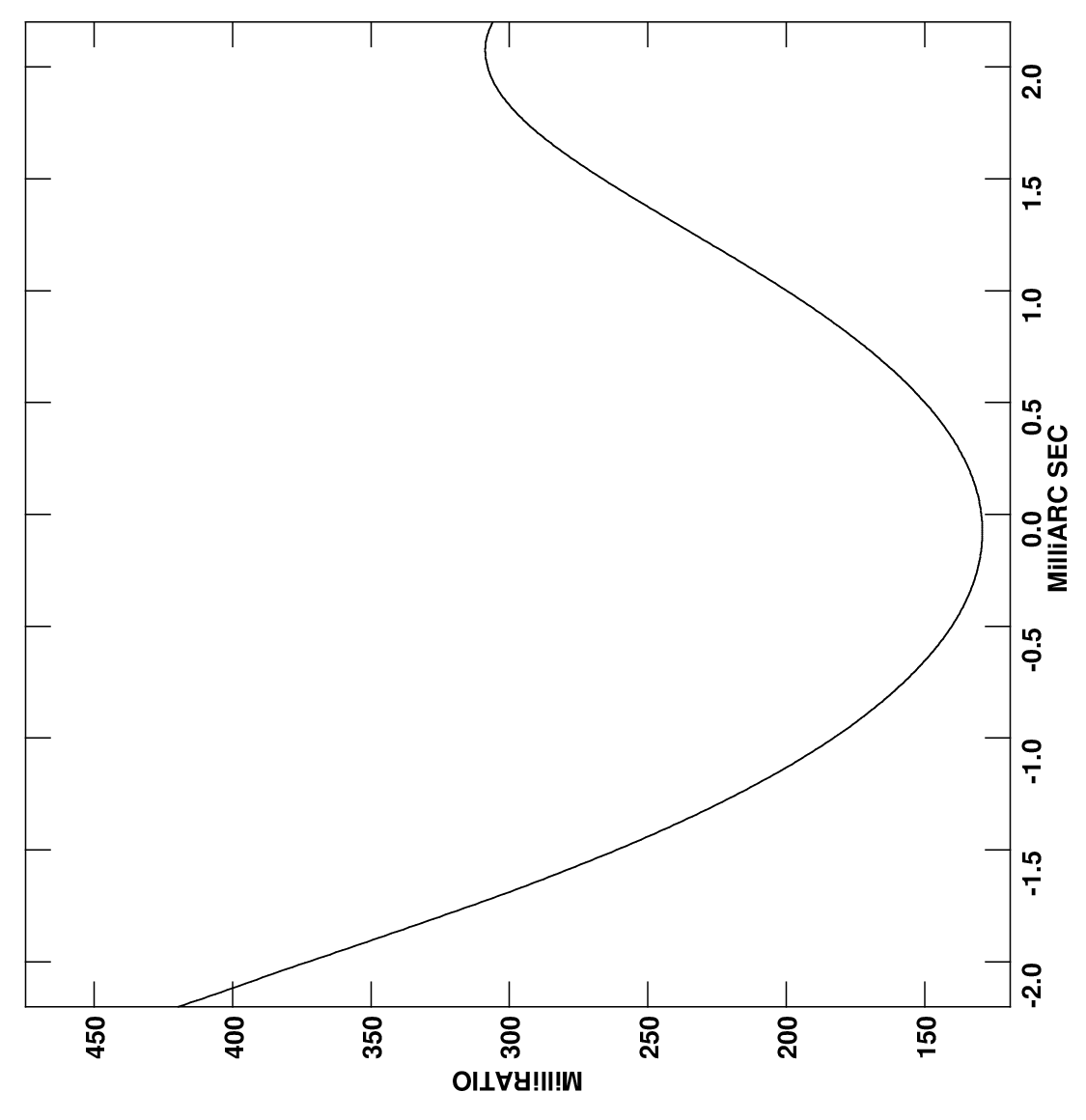}
\caption{The 5 GHz $m$ profiles across the jet on slices: (a) in the
bottom region with P.A. of $115^\circ$, (b) in the middle region
with P.A. of $54^\circ$, (c) in the top region with P.A. of
$118^\circ$, as indicated in Figure 1. \label{fig5}}
\end{figure*}

One may notice that in Figure 1,  the fractional polarization at 5
Ghz is relatively low along the local jet spine, but gets higher
toward the jet edges beyond $\sim$5 mas from the core region
(referred to as spine-sheath like structure hereafter). At 8 GHz,
although it is not so obvious, one can still find such a signature
at about 5 and 25 mas from the core, the only two regions with
polarization information available, while at 15, 22, and 43 GHz, it
is difficult to make sure that this feature exists for the current
data. To have quantitative idea on polarization level across the
jet, three slices perpendicular to local jet direction are selected
for $m$ profiles, with the local jet direction defined to be that
pointing from one component to the next down the jet at 5 GHz. The
fitted component distribution at 5 GHz is shown in
Figure~\ref{fig4}, with each component indicated as an elliptical
form by the fitted position, size and orientation.
One can find that the radio structure is well reproduced with these
components. In the core region containing 4 fitted components, we
cannot definitely determine the local jet direction according to the
component distribution, and choose the component's moving direction
with positional angle (P.A.) of $25^\circ$ as the local jet
direction from \citet{lis09}. The selected slices lie in 3 separate
regions of detectable polarization emission with  P.A. of
$115^\circ$, $54^\circ$ and $118^\circ$ from south to north, as is
indicated in Figure 1. The $m$ profiles on these slices are shown in
Figure~\ref{fig5}, displaying that the polarization level is
relatively low in the jet spine, but gets higher towards both sides
of the jet.
If we ignore the patch of extremely high fractional polarization at
the right bottom of the middle region, this trend would appear a bit
more remarkable.

\subsection{Projected polarization structure and Faraday rotation}

The projected polarization structure throughout the 5 frequencies
are shown in the right panel of each map in Figures \ref{fig1} and
\ref{fig2}, where the linearly polarized intensity distribution is
superposed. As can be seen in all these images, the electric field
is locally well-ordered from the core to the outmost emission
region. For the 5 GHz polarization image with high signal to noise
ratio, the electric field information can be well extracted on a
scale of more than 30 mas from the core, corresponding to a linear
size of $240$ pc. At this frequency, the electric vector is
distributed with P.A. of about $50^\circ$ near the core, and then
bifurcate apparently at a distance of about $5$ mas north to the
core (see also the polarization image at 8 GHz). When the jet
reaches a distance of about $10$ mas from the core, the electric
field turns almost perpendicular to the overall jet direction with
an overall P.A of \textbf{$\sim78^\circ$}. At a distance of about
$25$ mas north to the core, the electric vector becomes roughly
parallel to the north direction with an overall P.A. of
\textbf{$\sim23^\circ$}. Such an orientation change down the jet may
be attributed to magnetic fields that are ordered by local phenomena
at various places in the jet, while \citet{gab04} suggested that
these alternating magnetic fields may indicate oscillations or
instabilities of a global jet magnetic field.

\begin{figure}
\includegraphics[angle= -90,width=8cm, totalheight=7cm]{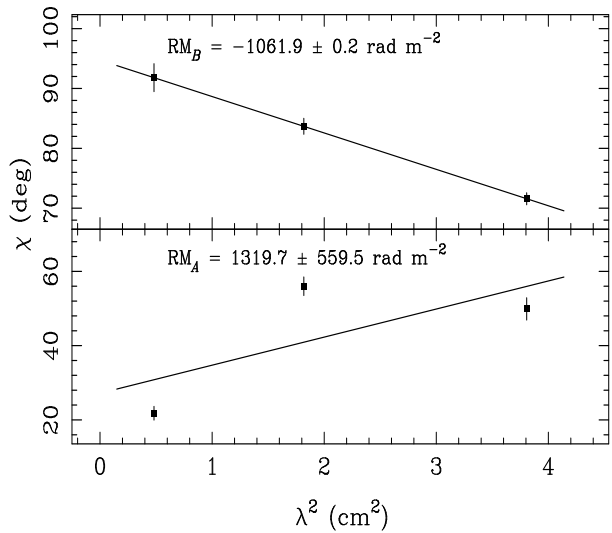}
\caption{A $\lambda^2$ regression to the observed Faraday rotation
 as a solid line for component \emph{A} (the bottom panel) and
 component \emph{B} (the top panel). The uncertainties of RMs are derived assuming the
 reduced chi-square value of unity.\label{fig6}}
\end{figure}

Faraday rotation reveals some physical condition along the line of
sight. When the polarized emission propagates through a magnetized
plasma, the polarization plane will rotates with wavelength
$\lambda$ according to the following expression (e.g.
\citealt{tay98}),
\begin{equation}\label{eq2}
    \Delta\chi=812\lambda^2\int
    \rm{N(\textbf{s})}\textbf{B(s)}\cdot d\textbf{s}=RM\lambda^2,
\end{equation}
where $\Delta\chi$ is polarization angle offset due to Faraday
rotation, $\rm{N(\textbf{s})}$ is the electron number density in
$\rm{cm^{-3}}$, \textbf{B(s)} is magnetic field in mG, and the
integral is taken over a passage \textbf{s} in parsec along the line
of sight from the source to the observer, RM is the rotation measure
in $\rm{rad~m^{-2}}$. The RM distribution can be used to explore the
intervening plasma and magnetic field along the line of sight. To
ensure proper estimation of RM, at least 3 band of EVPAs are
required. Under the limitation of resolution and sensitivity for the
current observational data, we cannot obtain a transverse RM
variation across several beamwidths. Based on the resultant EVPAs
between 15 and 43 GHz from model fitting, the components \emph{A}
and \emph{B} are fitted with equation (2). As a result, the
component \emph{A} is badly fitted, while the EVPAs for the
component \emph{B} obeyed a $\lambda^2$ Faraday rotation law well,
with the observed RM of $-1061.9 \pm 0.2~ \rm{rad~m^{-2}}$. The
fitting result for both components is shown in Figure~\ref{fig6},
where the errors are estimated from the Stokes $Q$ and $U$ images
with $5\sigma$ of noise level adopted.
In view of that the absolute RM due to our galaxy is usually no more
than $200~ \rm{rad~m^{-2}}$ in any direction \citep{pus01}, and the
RM in the rest frame of \mbox{NRAO 530} are higher by a factor of
$(1 + z)^2$, the RM in its rest frame will become quite large, with
the absolute value of more than 3000 $\rm{rad~m^{-2}}$, which
implies that most of the RM should arise from the source itself or
plasma near the source, rather than the plasma in our galaxy.

Since there exists quite a large Faraday rotation for component
\emph{B}, it naturally reminds one that for the Faraday rotation
occurring within the jet, radiation emitted at different depths is
rotated through different angles, thus the net flux will be
depolarized, and the polarization level will fall correspondingly.
By contrast, external Faraday rotation is relatively hard to result
in depolarization (e.g. \citealt{bur66}). For an optically-thin
uniform source, the internal Faraday dispersion can be described by
an additional factor of $\rm{sin(\Phi)/\Phi}$ where $\Phi =
2.0\Delta\chi$ \citep{bur66, hom09}. Based on the equation
(\ref{eq2}), the additional factors are obtained with values of
0.895, 0.975, and 0.998 at the corresponding frequency 15, 22, and
43 GHz, respectively. This means that even if all the observed
rotation is completely internal to the jet, this rotation is not
large enough to cause all of the decrease in fractional polarization
observed, assuming that the component has comparable degree of
polarization without depolarization. \citet{hom09} argued that the
alternative cause inducing depolarization is the opacity by the
mediums in the passage, by which the depolarization also gets worse
with wavelength. This implies that both the opacity and internal
Faraday rotation contribute to the decrease in fractional
polarization with wavelength, in which the opacity plays quite a
large part of role in the depolarization.

\section{DISCUSSION}

\subsection{Core identification}
In general, the core component in blazars lies in the extreme end of
the jet, which is usually optically thick, and highly Doppler
boosted with dominant flux density, relatively hard spectrum and
high brightness temperature. By comparing the physical quantities of
component \emph{A} and \emph{B} in Table~\ref{tb2}, one can find
that component \emph{A} has higher brightness temperature and harder
spectrum, both of which are indicators of radio core (e.g.
\citealt{she05}). On the other hand, the component \emph{B} exhibits
higher flux density and lower fractional polarization, which are
also taken as the features of a radio core (e.g.
\citealt{bow97,cot97}). These apparently contrary observational
evidences for core identification exists in both of the components,
which increases the difficulty in core identification.

Due to large Doppler boosting effect present in \mbox{NRAO 530}
\citep{bow97,jos01}, it is impossible for the jet structure to be
two-sided. Therefore, regardless of which one is the core component,
the jet direction changes down the jet to connect the
north-northwest structure \citep{jos01}. This will lead to changes
of Doppler boosting effect down the jet. Assuming that the component
\emph{A} is the core of the object, the adverse condition of
relatively high fractional polarization and low flux density in core
identification may be explained as follows: the component \emph{B}
has a relatively high flux density mainly due to the strong outburst
occurring two years before \citep{jos01}, and possibly larger
Doppler effect may also have contribution to it due to the change of
jet viewing angle, which require further confirmation; as for its
relatively high polarization level, it is because the component
\emph{A} itself consists at least two components (as is shown in the
radio structure of higher resolution at 86 GHz \citep{bow97}), of
which the jet component enhances the overall fractional polarization
of the fitted component \emph{A}. For the same reason, the component
\emph{A} cannot be fitted well with a single power low spectrum (see
Figure~\ref{fig3}), and the EVPAs doesn't obey the wavelength square
law as well (see Table~\ref{tb2}) \citep{tay98}. If, on the
contrary, the component \emph{B} is the core, it is not only hard to
explain its relatively soft spectrum, low brightness temperature,
but also the trajectories are required to bend by nearly $180^\circ$
to connect the north-northwest structure, which would be quite
unnatural to interpret it logically. Combine all the above reasons,
plus the component \emph{A} lying in the south end of its radio
structure, we conclude that it is the component \emph{A} that is the
true core of the object, which is consistent with the argument
suggested in \citet{jos01} and \citet{fen06}.

\subsection{Magnetic geometry in jet}

As mentioned earlier, the magnetic geometry is crucial in
determining what role the magnetic field plays in the dynamics and
emission of relativistic jets in active galactic nuclei, which may
be helical due to the differential rotation of central engine (e.g.
\citealt{mei01}), transverse due to shock compression, longitudinal
due to the effect of shear or interaction with external medium (e.g.
\citealt{lai80}), or chaotic. Through polarimetric observations, the
helical magnetic fields may manifest as spine-sheath polarization
structure (the polarization electric vectors predominantly aligned
with the jet in the jet spine and perpendicular to the jet at one or
both edges) (e.g. \citealt{pus05}), {fractional polarization
relatively low in the jet spine and increasing towards the jet edges
\citep{lyu05}), and/or the exclusive feature of RM gradient across
the jet \citep{bla93,asa02}.

\subsubsection{Interpretation with magnetic compression and shear}

As mentioned earlier, NRAO 530 shows the transverse $m$ profile
relatively low along the jet spine, and progressively increasing
toward the jet edges, suggesting that the integrated magnetic field
along the line of sight becomes better ordered toward the jet edges
\citep{gom08}. One possibility for presence of spine-sheath like
structure as well as the magnetic bifurcate configuration might be
related to the combination of shocks and flow shears
\citep{lai80,lai06}. The shocks compress and partially order an
initially tangled magnetic field, and the flow shear across the jet
stretches the magnetic field along the jet \citep{war94}. The
combination of shock compressions and flow velocity gradient across
the jet may well reproduce the magnetic bifurcate configuration and
the spine-sheath like $m$ profiles (e.g. \citealt{lai80,lai06}),
which is found in the $m$ and polarization structure of \mbox{NRAO
530}.

Another important observational feature for magnetic shock
compression in AGNs is that the magnetic field projected on the sky
plane is transverse to the local jet axis. This is because the
relativistic shocks enhance the magnetic component in the plane of
compression, perpendicular to the direction of propagation of the
shock \citep{lai80,hug89}. From the polarization structures at 5 and
8 GHz, the overall magnetic field directions in the bottom and top
regions are roughly perpendicular to the local jet directions, which
agrees well with the shock compression model. Furthermore, for the
top two regions with detectable linear polarization at 5 GHz, one
can find that the largest variation in fractional polarization
appears to be North to South, where the polarization goes to zero
between the two regions. The highest levels of fractional
polarization are at the top (north) and bottom (south). It is likely
that these two regions belongs to a bigger bright shocked one, where
the roughly transverse shocked field has canceled a roughly
longitudinal field in the underlying jet. This will produce the
effect that the middle of the structure appears to have the lowest
polarization due to cancelation, as is predicted by shock in jet
models (e.g. \citealt{gab01,hom09}). Similar cases were also
reported in some other sources (e.g. \citealt{gab01}) with
alternating aligned and orthogonal polarization vectors down the
jet, where the longitudinal magnetic field mainly ascribed to the
flow shear, oblique or conical shocks.

\subsubsection{Helical interpretation of the magnetic configuration}

It is also possible that the emission occurs in a large scale of
helical magnetic field with certain combinations of pitch and
viewing angle, resulting in the spine-sheath like $m$ profiles in
NRAO 530. For example, similar $m$ profiles across the jet are
presented with the viewing angle of $1/\Gamma$ or $1/2\Gamma$ in the
observer's frame and pitch angle of $45^\circ$ in the rest frame
shown in Figure 9 in \citet{lyu05} (here $\Gamma$ is Lorentz
factor). The extremely high fractional polarization at 5 and 8 GHz
in the outmost emission region from the core might also be a
manifestation of such a structure.

In addition, about 5 mas away from the core at 5 and 8 Ghz, the
polarization vectors bifurcate obviously to opposite sides from the
local jet axis, something like a spine-sheath structure. At
distances of 12 and 25 mas from the core at 5 GHz, such a structure
can also roughly be seen, though it is not very obvious. Synchrotron
emission of relativistic particles in a helical magnetic field can
naturally explain the polarization structure in the jet frame (e.g.
\citealt{pus05}).

\vspace{3mm}One may have noticed that the polarimetric properties of
the object available for the current observations can not
exclusively make it clear the magnetic configuration over the whole
structure. Through multi-bands polarimetric VLBI observations of
\mbox{3C 279} with the same 3 fitted components available at all the
6 bands from 8 to 22 GHz, \citet{hom09} obtained a self-consistent
picture for its magnetic configuration, particle population and low
cutoff energy range by numerical simulations to the full
polarization spectra, which cannot be achieved in our case due
mainly to the limitation of the observational data and partly to the
source structure itself. Additionally, at the 15 GHz and higher
frequencies, the polarization structure is not resolvable over the
jet width, which prevent us getting the transverse RM variation down
the jet for further analysis on the magnetic configuration. It is
noticeable that large orientation changes of polarization structure
down the jet have also been detected in PKS 1418+546 and OJ287
\citep{gab01, gab03}. They both show a spine-sheath like structure
in fractional polarization at the same time. This probably implies
that there exists a certain connection between spine-sheath like
structure and large orientation change of electric vectors down the
jet, which is worthy of further investigation.

\section{SUMMARY}
We have performed the polarization-sensitive VLBI observations at 5,
8, 15, 22, and 43 GHz within a week in February 1997. Total
intensity, fractional polarization, linear intensity and EVPA
distributions at all these frequencies are presented. Model fitting
has been done to the full polarimetric visibility data at all the 5
frequencies, with focus mainly on the two components \emph{A} and
\emph{B} in the core region. Comparing to the component \emph{B},
the southmost component \emph{A} shows relatively high brightness
temperature, hard spectrum, which is identified as the radio core of
the object. The relatively high polarization level for the component
is probably because it contains an additional jet component of high
fractional polarization, which is resolved at 86 GHz with a VLBI
observation of higher resolution \citep{bow97}. Whereas the
component \emph{B} of dominant flux density exhibits a good
power-law spectrum with steady increase in fractional polarization
with frequency from 15 to 43 GHz.  The observed EVPAs for the
component \emph{B} is in agreement well with the $\lambda^2$ law,
with the observed RM of about $-1062~\rm{rad~m^{-2}}$. Assuming that
the component has a comparable degree of polarization without
depolarization at these frequencies, both the opacity and internal
Faraday rotation have effects on the decrease in fractional
polarization with wavelength, in which the opacity plays quite a
large part of role.

The linear polarization shows a spine-sheath like structure in some
regions at 5 and 8 GHz with degree of polarization relatively low
along the jet spine, but getting higher towards both edges. The
largest variation in fractional polarization appears to be north to
south, where the polarization goes to zero between the top two
regions. The highest levels of fractional polarization occur at the
bottom and top, while the lowest one occurs in the middle. The
polarization structure at 5 GHz shows that the magnetic fields
appear alternately orthogonal and aligned down the jet, with a
signature of bifurcating to the opposite sides from the local jet
spine. All these radiative features can be explained either with a
large scale of helical magnetic field present within the jet, or
with tangled magnetic fields compressed and sheared down the jet.
Further polarimetric VLBI observations are required with sufficient
high resolution and sensitivity at multiple wavelengths to further
determine if the magnetic field is helical or not.

\section*{Acknowledgments}

The authors do appreciate the anonymous referee for insightful
comments and constructive suggestions, which were greatly helpful in
improving our paper, we also thank Prof. Jiang for helpful
discussions. This work was supported by the National Natural Science
Foundation of China (grants 10573029, 10625314, 10633010 and
10821302) and the Knowledge Innovation Program of the Chinese
Academy of Sciences (Grant No. KJCX2-YW-T03), by the Science and
Technology Commission of Shanghai Municipality (09ZR1437400) and the
Scientific Research Foundation for Returned Scholars, Ministry of
Education of China (9020090306),  by the Program of Shanghai Subject
Chief Scientist (06XD14024) and the National Key Basic Research
Development Program of China (No. 2007CB815405). This paper made use
of data from the University of Michigan Radio Astronomy Observatory,
supported by the University of Michigan and the National Science
Foundation. ZQS acknowledges the support by the One-Hundred-Talent
Program of Chinese Academy of Sciences.

\label{lastpage}

\end{document}